# The Transverse Momentum and Total Cross Section of $e^+e^-$ Pairs in the Z-boson Region from $p\bar{p}$ Collisions at $\sqrt{s}$ = 1.8 TeV

T. Affolder et al.
The CDF Collaboration

*Fermi National Accelerator Laboratory*
*P.O. Box 500, Batavia, Illinois 60510*

# The Transverse Momentum and Total Cross Section of $e^+e^-$ Pairs in the $Z$-boson Region from $p\bar{p}$ Collisions at $\sqrt{s} = 1.8$ TeV

T. Affolder,[21] H. Akimoto,[42] A. Akopian,[35] M. G. Albrow,[10] P. Amaral,[7] S. R. Amendolia,[31] D. Amidei,[24] J. Antos,[1] G. Apollinari,[35] T. Arisawa,[42] T. Asakawa,[40] W. Ashmanskas,[7] M. Atac,[10] P. Azzi-Bacchetta,[29] N. Bacchetta,[29] M. W. Bailey,[26] S. Bailey,[14] P. de Barbaro,[34] A. Barbaro-Galtieri,[21] V. E. Barnes,[33] B. A. Barnett,[17] M. Barone,[12] G. Bauer,[22] F. Bedeschi,[31] S. Belforte,[39] G. Bellettini,[31] J. Bellinger,[43] D. Benjamin,[9] J. Bensinger,[4] A. Beretvas,[10] J. P. Berge,[10] J. Berryhill,[7] S. Bertolucci,[12] B. Bevensee,[30] A. Bhatti,[35] C. Bigongiari,[31] M. Binkley,[10] D. Bisello,[29] R. E. Blair,[2] C. Blocker,[4] K. Bloom,[24] S. Blusk,[34] A. Bocci,[31] A. Bodek,[34] W. Bokhari,[30] G. Bolla,[33] Y. Bonushkin,[5] D. Bortoletto,[33] J. Boudreau,[32] A. Brandl,[26] S. van den Brink,[17] C. Bromberg,[25] N. Bruner,[26] E. Buckley-Geer,[10] J. Budagov,[8] H. S. Budd,[34] K. Burkett,[14] G. Busetto,[29] A. Byon-Wagner,[10] K. L. Byrum,[2] M. Campbell,[24] A. Caner,[31] W. Carithers,[21] J. Carlson,[24] D. Carlsmith,[43] J. Cassada,[34] A. Castro,[29] D. Cauz,[39] A. Cerri,[31] P. S. Chang,[1] P. T. Chang,[1] J. Chapman,[24] C. Chen,[30] Y. C. Chen,[1] M. -T. Cheng,[1] M. Chertok,[37] G. Chiarelli,[31] I. Chirikov-Zorin,[8] G. Chlachidze,[8] F. Chlebana,[10] L. Christofek,[16] M. L. Chu,[1] S. Cihangir,[10] C. I. Ciobanu,[27] A. G. Clark,[13] M. Cobal,[31] E. Cocca,[31] A. Connolly,[21] J. Conway,[36] J. Cooper,[10] M. Cordelli,[12] J. Guimaraes da Costa,[24] D. Costanzo,[31] D. Cronin-Hennessy,[9] R. Cropp,[23] R. Culbertson,[7] D. Dagenhart,[41] F. DeJongh,[10] S. Dell'Agnello,[12] M. Dell'Orso,[31] R. Demina,[10] L. Demortier,[35] M. Deninno,[3] P. F. Derwent,[10] T. Devlin,[36] J. R. Dittmann,[10] S. Donati,[31] J. Done,[37] T. Dorigo,[14] N. Eddy,[16] K. Einsweiler,[21] J. E. Elias,[10] E. Engels, Jr.,[32] W. Erdmann,[10] D. Errede,[16] S. Errede,[16] Q. Fan,[34] R. G. Feild,[44] C. Ferretti,[31] I. Fiori,[3] B. Flaugher,[10] G. W. Foster,[10] M. Franklin,[14] J. Freeman,[10] J. Friedman,[22] Y. Fukui,[20] S. Gadomski,[23] S. Galeotti,[31] M. Gallinaro,[35] T. Gao,[30] M. Garcia-Sciveres,[21] A. F. Garfinkel,[33] P. Gatti,[29] C. Gay,[44] S. Geer,[10] D. W. Gerdes,[24] P. Giannetti,[31] P. Giromini,[12] V. Glagolev,[8] M. Gold,[26] J. Goldstein,[10] A. Gordon,[14] A. T. Goshaw,[9] Y. Gotra,[32] K. Goulianos,[35] H. Grassmann,[39] C. Green,[33] L. Groer,[36] C. Grosso-Pilcher,[7] M. Guenther,[33] G. Guillian,[24] R. S. Guo,[1] C. Haber,[21] E. Hafen,[22] S. R. Hahn,[10] C. Hall,[14] T. Handa,[15] R. Handler,[43] W. Hao,[38] F. Happacher,[12] K. Hara,[40] A. D. Hardman,[33] R. M. Harris,[10] F. Hartmann,[18] K. Hatakeyama,[35] J. Hauser,[5] J. Heinrich,[30] A. Heiss,[18] B. Hinrichsen,[23] K. D. Hoffman,[33] C. Holck,[30] R. Hollebeek,[30] L. Holloway,[16] R. Hughes,[27] J. Huston,[25] J. Huth,[14] H. Ikeda,[40] M. Incagli,[31] J. Incandela,[10] G. Introzzi,[31] J. Iwai,[42] Y. Iwata,[15] E. James,[24] H. Jensen,[10] M. Jones,[30] U. Joshi,[10] H. Kambara,[13] T. Kamon,[37] T. Kaneko,[40] K. Karr,[41] H. Kasha,[44] Y. Kato,[28] T. A. Keaffaber,[33] K. Kelley,[22] M. Kelly,[24] R. D. Kennedy,[10] R. Kephart,[10] D. Khazins,[9] T. Kikuchi,[40] M. Kirk,[4] B. J. Kim,[19] H. S. Kim,[23] S. H. Kim,[40] Y. K. Kim,[21] L. Kirsch,[4] S. Klimenko,[11] D. Knoblauch,[18] P. Koehn,[27] A. Köngeter,[18] K. Kondo,[42] J. Konigsberg,[11] K. Kordas,[23] A. Korytov,[11] E. Kovacs,[2] J. Kroll,[30] M. Kruse,[34] S. E. Kuhlmann,[2] K. Kurino,[15] T. Kuwabara,[40] A. T. Laasanen,[33] N. Lai,[7] S. Lami,[35] S. Lammel,[10] J. I. Lamoureux,[4] M. Lancaster,[21] G. Latino,[31] T. LeCompte,[2] A. M. Lee IV,[9] S. Leone,[31] J. D. Lewis,[10] M. Lindgren,[5] T. M. Liss,[16] J. B. Liu,[34] Y. C. Liu,[1] N. Lockyer,[30] M. Loreti,[29] D. Lucchesi,[29] P. Lukens,[10] S. Lusin,[43] J. Lys,[21] R. Madrak,[14] K. Maeshima,[10] P. Maksimovic,[14] L. Malferrari,[3] M. Mangano,[31] M. Mariotti,[29] G. Martignon,[29] A. Martin,[44] J. A. J. Matthews,[26] P. Mazzanti,[3] K. S. McFarland,[34] P. McIntyre,[37] E. McKigney,[30] M. Menguzzato,[29] A. Menzione,[31] E. Meschi,[31] C. Mesropian,[35] C. Miao,[24] T. Miao,[10] R. Miller,[25] J. S. Miller,[24] H. Minato,[40] S. Miscetti,[12] M. Mishina,[20] N. Moggi,[31] E. Moore,[26] R. Moore,[24] Y. Morita,[20] A. Mukherjee,[10] T. Muller,[18] A. Munar,[31] P. Murat,[31] S. Murgia,[25] M. Musy,[39] J. Nachtman,[5] S. Nahn,[44] H. Nakada,[40] T. Nakaya,[7] I. Nakano,[15] C. Nelson,[10] D. Neuberger,[18] C. Newman-Holmes,[10] C.-Y. P. Ngan,[22] P. Nicolaidi,[39] H. Niu,[4] L. Nodulman,[2] A. Nomerotski,[11] S. H. Oh,[9] T. Ohmoto,[15] T. Ohsugi,[15] R. Oishi,[40] T. Okusawa,[28] J. Olsen,[43] C. Pagliarone,[31] F. Palmonari,[31] R. Paoletti,[31] V. Papadimitriou,[38] S. P. Pappas,[44] A. Parri,[12] D. Partos,[4] J. Patrick,[10] G. Pauletta,[39] M. Paulini,[21] A. Perazzo,[31] L. Pescara,[29] T. J. Phillips,[9] G. Piacentino,[31] K. T. Pitts,[10] R. Plunkett,[10] A. Pompos,[33] L. Pondrom,[43] G. Pope,[32] F. Prokoshin,[8] J. Proudfoot,[2] F. Ptohos,[12] G. Punzi,[31] K. Ragan,[23] D. Reher,[21] A. Ribon,[29] F. Rimondi,[3] L. Ristori,[31] W. J. Robertson,[9] A. Robinson,[23] T. Rodrigo,[6] S. Rolli,[41] L. Rosenson,[22] R. Roser,[10] R. Rossin,[29] W. K. Sakumoto,[34] D. Saltzberg,[5] A. Sansoni,[12] L. Santi,[39] H. Sato,[40] P. Savard,[23] P. Schlabach,[10] E. E. Schmidt,[10] M. P. Schmidt,[44] M. Schmitt,[14] L. Scodellaro,[29] A. Scott,[5] A. Scribano,[31] S. Segler,[10] S. Seidel,[26] Y. Seiya,[40] A. Semenov,[8] F. Semeria,[3] T. Shah,[22] M. D. Shapiro,[21] P. F. Shepard,[32] T. Shibayama,[40] M. Shimojima,[40] M. Shochet,[7] J. Siegrist,[21] G. Signorelli,[31] A. Sill,[38] P. Sinervo,[23] P. Singh,[16] A. J. Slaughter,[44] K. Sliwa,[41] C. Smith,[17] F. D. Snider,[10] A. Solodsky,[35] J. Spalding,[10] T. Speer,[13] P. Sphicas,[22] F. Spinella,[31] M. Spiropulu,[14] L. Spiegel,[10] L. Stanco,[29] J. Steele,[43] A. Stefanini,[31] J. Strologas,[16] F. Strumia, [13] D. Stuart,[10] K. Sumorok,[22] T. Suzuki,[40] R. Takashima,[15] K. Takikawa,[40] M. Tanaka,[40] T. Takano,[28] B. Tannenbaum,[5] W. Taylor,[23] M. Tecchio,[24] P. K. Teng,[1] K. Terashi,[40] S. Tether,[22] D. Theriot,[10] R. Thurman-Keup,[2] P. Tipton,[34] S. Tkaczyk,[10] K. Tollefson,[34] A. Tollestrup,[10] H. Toyoda,[28] W. Trischuk,[23] J. F. de Troconiz,[14] S. Truitt,[24] J. Tseng,[22] N. Turini,[31] F. Ukegawa,[40] J. Valls,[36] S. Vejcik III,[10] G. Velev,[31] R. Vidal,[10] R. Vilar,[6] I. Vologouev,[21] D. Vucinic,[22] R. G. Wagner,[2] R. L. Wagner,[10] J. Wahl,[7] N. B. Wallace,[36]

A. M. Walsh,[36] C. Wang,[9] C. H. Wang,[1] M. J. Wang,[1] T. Watanabe,[40] T. Watts,[36] R. Webb,[37] H. Wenzel,[18] W. C. Wester III,[10] A. B. Wicklund,[2] E. Wicklund,[10] H. H. Williams,[30] P. Wilson,[10] B. L. Winer,[27] D. Winn,[24] S. Wolbers,[10] D. Wolinski,[24] J. Wolinski,[25] S. Worm,[26] X. Wu,[13] J. Wyss,[31] A. Yagil,[10] W. Yao,[21] G. P. Yeh,[10] P. Yeh,[1] J. Yoh,[10] C. Yosef,[25] T. Yoshida,[28] I. Yu,[19] S. Yu,[30] A. Zanetti,[39] F. Zetti,[21] and S. Zucchelli[3]

(CDF Collaboration)

[1] *Institute of Physics, Academia Sinica, Taipei, Taiwan 11529, Republic of China*
[2] *Argonne National Laboratory, Argonne, Illinois 60439*
[3] *Istituto Nazionale di Fisica Nucleare, University of Bologna, I-40127 Bologna, Italy*
[4] *Brandeis University, Waltham, Massachusetts 02254*
[5] *University of California at Los Angeles, Los Angeles, California 90024*
[6] *Instituto de Fisica de Cantabria, University of Cantabria, 39005 Santander, Spain*
[7] *Enrico Fermi Institute, University of Chicago, Chicago, Illinois 60637*
[8] *Joint Institute for Nuclear Research, RU-141980 Dubna, Russia*
[9] *Duke University, Durham, North Carolina 27708*
[10] *Fermi National Accelerator Laboratory, Batavia, Illinois 60510*
[11] *University of Florida, Gainesville, Florida 32611*
[12] *Laboratori Nazionali di Frascati, Istituto Nazionale di Fisica Nucleare, I-00044 Frascati, Italy*
[13] *University of Geneva, CH-1211 Geneva 4, Switzerland*
[14] *Harvard University, Cambridge, Massachusetts 02138*
[15] *Hiroshima University, Higashi-Hiroshima 724, Japan*
[16] *University of Illinois, Urbana, Illinois 61801*
[17] *The Johns Hopkins University, Baltimore, Maryland 21218*
[18] *Institut für Experimentelle Kernphysik, Universität Karlsruhe, 76128 Karlsruhe, Germany*
[19] *Korean Hadron Collider Laboratory: Kyungpook National University, Taegu 702-701; Seoul National University, Seoul 151-742; and SungKyunKwan University, Suwon 440-746; Korea*
[20] *High Energy Accelerator Research Organization (KEK), Tsukuba, Ibaraki 305, Japan*
[21] *Ernest Orlando Lawrence Berkeley National Laboratory, Berkeley, California 94720*
[22] *Massachusetts Institute of Technology, Cambridge, Massachusetts 02139*
[23] *Institute of Particle Physics: McGill University, Montreal H3A 2T8; and University of Toronto, Toronto M5S 1A7; Canada*
[24] *University of Michigan, Ann Arbor, Michigan 48109*
[25] *Michigan State University, East Lansing, Michigan 48824*
[26] *University of New Mexico, Albuquerque, New Mexico 87131*
[27] *The Ohio State University, Columbus, Ohio 43210*
[28] *Osaka City University, Osaka 588, Japan*
[29] *Universita di Padova, Istituto Nazionale di Fisica Nucleare, Sezione di Padova, I-35131 Padova, Italy*
[30] *University of Pennsylvania, Philadelphia, Pennsylvania 19104*
[31] *Istituto Nazionale di Fisica Nucleare, University and Scuola Normale Superiore of Pisa, I-56100 Pisa, Italy*
[32] *University of Pittsburgh, Pittsburgh, Pennsylvania 15260*
[33] *Purdue University, West Lafayette, Indiana 47907*
[34] *University of Rochester, Rochester, New York 14627*
[35] *Rockefeller University, New York, New York 10021*
[36] *Rutgers University, Piscataway, New Jersey 08855*
[37] *Texas A&M University, College Station, Texas 77843*
[38] *Texas Tech University, Lubbock, Texas 79409*
[39] *Istituto Nazionale di Fisica Nucleare, University of Trieste/ Udine, Italy*
[40] *University of Tsukuba, Tsukuba, Ibaraki 305, Japan*
[41] *Tufts University, Medford, Massachusetts 02155*
[42] *Waseda University, Tokyo 169, Japan*
[43] *University of Wisconsin, Madison, Wisconsin 53706*
[44] *Yale University, New Haven, Connecticut 06520*

The transverse momentum and total cross section of $e^+e^-$ pairs in the $Z$-boson region of $66 < M_{ee} < 116$ GeV/$c^2$ from $p\bar{p}$ collisions at $\sqrt{s} = 1.8$ TeV are measured using 110 pb$^{-1}$ of collisions taken by the Collider Detector at Fermilab during 1992–1995. The total cross section is measured to be $248 \pm 11$ pb. The differential transverse momentum cross section is compared with calculations that match quantum chromodynamics perturbation theory at high transverse momentum with the gluon resummation formalism at low transverse momentum.

In hadron-hadron collisions at high energies, massive $e^+e^-$ pairs are produced via the Drell-Yan [1] process. In the standard model, colliding partons from the hadrons react to form an intermediate $\gamma^*$ or $Z$ ($\gamma^*/Z$) vector boson, which then decays into an $e^+e^-$ pair. Initial state quantum chromodynamic (QCD) radiation from the colliding partons give the Drell-Yan pair a transverse momentum ($P_{\rm T}$) boost and a companion jet or jets. A study on the properties of high energy jets produced with $e^+e^-$ pairs from $Z$ bosons [2] finds that QCD predictions agree with measurements. By measuring the production properties of $e^+e^-$ pairs, QCD can be studied in even more detail because state-of-the-art predictions for them are finite and physical at all $P_{\rm T}$. The calculations contain fixed-order perturbation theory at high $P_{\rm T}$ and gluon resummation formalisms [3,4], which sum perturbative contributions from multiple soft and collinear gluon emissions, at low $P_{\rm T}$. While calculations and previous measurements in $p\bar{p}$ collisions at $\sqrt{s} = 0.63$ TeV [5] and $\sqrt{s} = 1.8$ TeV [6–8] are consistent, the physics at low $P_{\rm T}$ is data driven because nonperturbative physics effects are also important. These effects are postulated to factorize into a form factor, which, like a parton distribution function, is expected to be universal. As QCD cannot yet be solved exactly, this physics can only be obtained by measurements.

In this Letter, precise new measurements of the differential $P_{\rm T}$ and total cross sections of $e^+e^-$ pairs in the mass range 66–116 GeV/$c^2$ are presented. They are denoted as $d\sigma/dP_{\rm T}$ and $\sigma$, respectively, and are corrected for acceptance, efficiencies, and detector resolution effects. The $e^+e^-$ pairs in this mass range are mostly from the resonant production and decay of $Z$ bosons.

The $e^+e^-$ pairs are from 110 pb$^{-1}$ of $p\bar{p}$ collisions at $\sqrt{s} = 1.8$ TeV taken by the Collider Detector at Fermilab [9] (CDF) during 1992–1993 ($19.7 \pm 0.7$ pb$^{-1}$) and 1994–1995 ($90.4 \pm 3.7$ pb$^{-1}$). CDF is a solenoidal magnetic spectrometer surrounded by projective-tower-geometry calorimeters and outer muon detectors. Only detector components used in this measurement are described here. Charged particle momenta and directions are measured by the spectrometer, which consists of a 1.4 T axial magnetic field, an 84-layer cylindrical drift chamber (CTC), and an inner vertex tracking chamber (VTX). The polar coverage of the CTC tracking is $|\eta| < 1.2$. The $p\bar{p}$ collision point along the beam line ($Z_{\rm vertex}$) is determined using tracks in the VTX. The $Z_{\rm vertex}$ distribution is approximately Gaussian, with $\sigma \sim 26$ cm, and nominally centered in the middle of CDF ($z = 0$). The energies and directions of electrons, photons, and jets are measured by the calorimeters covering three regions: central ($|\eta| < 1.1$), end plug ($1.1 < |\eta| < 2.4$), and forward ($2.2 < |\eta| < 4.2$). Each region has an electromagnetic (EM) and hadronic (HAD) calorimeter. The central EM calorimeter's shower maximum strip detector (CES) is used for EM shower position measurements.

High mass Drell-Yan events are distinctive: the $e^+$ and $e^-$ typically have large transverse energies ($E_{\rm T}$), are separated from each other, and tend to be separated from jets and other particles from the interaction. The sample of Drell-Yan events used in this measurement was collected by a three-level online trigger [10] for $p\bar{p}$ collisions with a high $E_{\rm T}$ electron in the central calorimeter region. In the offline analysis, events with two or more electron candidates are selected. The $p\bar{p}$ vertex is required to be within the fiducial region $|Z_{\rm vertex}| < 60$ cm. This reduces the integrated luminosity by $(4.5 \pm 1.1)\%$ for the 1992–1993 data and $(6.3 \pm 1.1)\%$ for the 1994–1995 data. Since the electrons from the Drell-Yan process are typically isolated, both electrons are required to be isolated from any other activity in the calorimeters. Both electrons are also required to be within the fiducial area of the calorimeters. One electron must be within the central region, and it or another central electron must satisfy the trigger requirements. Electrons in the central, end plug, and forward regions are required to have a minimum $E_{\rm T}$ of 20, 15, and 15 GeV, respectively. There are three categories of $ee$ pairs: central-central (CC), central-end plug (CP), and central-forward (CF). Only $ee$ pairs in the mass range 66–116 GeV/$c^2$ are used.

To improve the purity of the sample, electron identification cuts are applied. Electron identification in the central region is the most powerful due to the additional discriminating power of tracking and the CES. The central electron (or one of them if there are two) is required to pass strict criteria [11]. The criteria on the other electron are looser. A central electron must have a CTC track that extrapolates onto the electron's shower clusters in the EM calorimeter and the CES. These clusters must have EM-like transverse shower profiles. The track momentum and the EM shower energy must be consistent with one another (i.e., $E/P \sim 1$). The track is also used to set the position and direction of the electron. The fraction of energy in the HAD calorimeter towers behind the EM shower is required to be consistent with that expected for an EM shower (the $E_{\rm had}/E_{\rm EM}$ requirement). End plug region electrons are required to have an EM-like transverse shower profile and to pass a loose $E_{\rm had}/E_{\rm EM}$ requirement. Forward region electrons are only required to pass a loose $E_{\rm had}/E_{\rm EM}$ requirement.

After all cuts, the numbers of CC, CP, and CF topology events are 2951, 3824, and 745, respectively. The backgrounds are small, and are estimated using the data. The backgrounds in the CP and CF topologies are dominated by jets and are $78 \pm 65$ and $27 \pm 13$ events, respectively. The background in the CC topology consists of $e^+e^-$ pairs from $W^+W^-$, $\tau^+\tau^-$, $c\bar{c}$, $b\bar{b}$, and $t\bar{t}$ sources. This background, estimated using oppositely charged $e\mu$ pairs [12], is $8 \pm 3$ events. Remaining jet backgrounds are negligible.

Since the mean instantaneous $p\bar{p}$ collision luminosity for the 1994–1995 run is about twice that for the 1992–1993 run, the efficiencies for each period are kept separate. The electron identification and isolation cut efficiencies of the 1992–1993 data for CC, CP, and CF pairs are $(79.6\pm1.8)\%$, $(80.5\pm1.4)\%$, and $(78.1\pm2.6)\%$, respec-tively. The corresponding efficiencies for the 1994–1995 data are $(78.3 \pm 0.9)\%$, $(78.0 \pm 0.7)\%$, and $(76.4 \pm 1.3)\%$. The mean trigger efficiency for one central electron is $(91.2 \pm 0.4)\%$ for the 1992–1993 data and $(89.6 \pm 0.2)\%$ for the 1994–1995 data. Most of the trigger inefficiency is due to the level-2 track finder, and the inefficiency from other trigger levels is less than 0.5%.

The measured cross sections are derived from $N = \sigma\mathcal{L}\cdot\epsilon A$, where $N$ is the number of background-subtracted events, $\sigma$ is the cross section, $\mathcal{L}$ is the integrated collision luminosity, $\epsilon$ is the event selection efficiency, and $A$ is the detector acceptance. The acceptance for Drell-Yan $e^+e^-$ pairs is obtained using the Monte Carlo event generator, PYTHIA [13], and CDF detector simulation programs. PYTHIA generates the hard, leading order (LO) QCD interaction, $q + \bar{q} \to \gamma^*/Z$, simulates initial state QCD radiation via its parton shower algorithms, and generates the decay, $\gamma^*/Z \to e^+e^-$. To approximate higher order QCD corrections to the LO mass distribution, a "$K$-factor" [14] is used as an event weight: $K(M^2) = 1 + \frac{4}{3}(1 + \frac{4}{3}\pi^2)\alpha_s(M^2)/2\pi$, where $\alpha_s$ is the two loop QCD coupling. This factor improves the agreement between the next-to-leading order (NLO) and LO mass spectra. For $M > 50$ GeV/$c^2$, $1.25 < K < 1.36$. PYTHIA is also used to generate the NLO QCD interactions, $q + \bar{q} \to \gamma^*/Z + g$, and $q(\bar{q}) + g \to \gamma^*/Z + q(\bar{q})$ for analysis at high $P_{\rm T}$. The CTEQ3L [15] nucleon parton distribution functions (PDFs) are used in the QCD calculations. Final state QED radiation [16] from the $\gamma^*/Z \to e^+e^-$ vertex is added by the PHOTOS [17] Monte Carlo. Generated events are processed by CDF detector simulation programs and then reconstructed as data.

The calorimetry energy scales and resolutions used in the detector simulation are tuned with data. The level-2 trigger is included in the detector simulation to accomodate a slight $E_{\rm T}$ and $\eta$ dependence of the efficiency. The simulated $e^+e^-$-pair $P_{\rm T}$ distribution is also adjusted [13] to agree with the data. After these adjustments, there is satisfactory agreement between the simulation and data for the $e^+e^-$-pair production and decay kinematics. Simulated events are accepted if, after the reconstruction, they pass the $e^+e^-$-pair mass cut, and the electron fiducial and $E_{\rm T}$ cuts. Thus, detector resolution effects are included in all acceptances.

For the $\sigma$ measurement, the acceptance is calculated using the LO QCD event generator. The overall acceptance of $e^+e^-$ pairs is $(37.6 \pm 0.1)\%$ for the 1992–1993 run and $(36.7 \pm 0.1)\%$ for the 1994–1995 run. The acceptances for the component CC, CP, and CF $e^+e^-$-pair topologies are 40, 50, and 10%, respectively, of the overall value. The CC, CP, and CF topology event samples provide three independent cross section measurements. Such measurements, derived from the 1992–1993 data, and separately for the 1994–1995 data, are consistent with each other. The combined measurement gives $\sigma = 248 \pm 4 \pm 3 \pm 10$ pb, where the first error contains the statistical and efficiency errors, the second error is the systematic error from background subtractions and the acceptance, and the last error is from the collision luminosity. The systematic errors are discussed later.

In the $d\sigma/dP_{\rm T}$ measurement, all the data are combined, binned in $P_{\rm T}$, and the cross section calculated with

$$\frac{d\sigma}{dP_{\rm T}} = \frac{\Delta N}{C\,\Delta P_{\rm T}\,\sum_r \mathcal{L}_r\,\epsilon A_r}.$$

The $\Delta N$ is the background-subtracted event count in a $P_{\rm T}$ bin, $C$ is a bin centering correction, $\Delta P_{\rm T}$ is the bin width, the sum $r$ is over the 1992–1993 and 1994–1995 runs, $\mathcal{L}_r$ is the run's integrated luminosity, and $\epsilon A_r$ is the run's combined event selection efficiency and acceptance. The $d\sigma/dP_{\rm T}$ is shown in Table I. The backgrounds subtracted from the event count are predicted using the data and background samples. The factor $C$ corrects the average value of the cross section in the bin to its bin center value. For the small bins of $P_{\rm T} < 50$ GeV/$c$, $1.00 < C < 1.01$, and for the larger bins above, $1.01 < C < 1.15$. In the acceptance calculation, PYTHIA's LO QCD event generator is used for the $P_{\rm T} < 20$ GeV/$c$ bins, and above that, its NLO QCD event generator is used. Separate acceptances are calculated for CC, CP, and CF topology $e^+e^-$ pairs. They are combined with the corresponding event selection efficiencies to give $\epsilon A_r$. As the $P_{\rm T}$ increases, jets opposing the $e^+e^-$ pair move further into the detector and cause acceptance losses via the electron isolation requirement. At $P_{\rm T} \sim 40$ GeV/$c$, the loss peaks then abates as both electrons are Lorentz boosted away from the jets. The relative loss, 8% at most, is taken from the simulation and included in $\epsilon A_r$. The correction $\epsilon A_r$ is a function of $P_{\rm T}$, and its minimum value is at $P_{\rm T} = 0$ GeV/$c$, where $\epsilon A_r \simeq 0.22$.

The systematic errors considered are from variations in the estimates of the total background, uncertainties in the calorimetry energy resolution functions used in the detector simulation, and variations of the Drell-Yan production model used by the Monte Carlo event generator. For the $\sigma$ measurement, the systematic errors from the background and Drell-Yan production model are both 1%. The calorimeter resolution effects are negligible. The detector resolution and Drell-Yan production model affect the detector's acceptance in $P_{\rm T}$. The estimated error from the calorimeter resolution is 3.6% at $P_{\rm T} = 0$ GeV/$c$, and it decreases to 0.6% for $P_{\rm T} > 20$ GeV/$c$. The estimated error from the Drell-Yan production model is under 3% for $P_{\rm T} < 30$ GeV/$c$ and 1% for higher $P_{\rm T}$. The $p\bar{p}$ collision luminosity is derived with CDF's beam-beam cross section, $\sigma_{\rm BBC} = 51.15 \pm 1.60$ mb [18,19]. The luminosity error contains the $\sigma_{\rm BBC}$ error and uncertainties specific to running conditions. For the 1994–1995 run, the luminosity error has recently been reduced to 4.1%.

The predictions for the production of $e^+e^-$ pairs in the mass range 66–116 GeV/$c^2$ are calculated using EV [3] and RESBOS [4]. These QCD calculations match perturbation theory at high $P_{\rm T}$ with gluon resummation formalisms at low $P_{\rm T}$. Nonperturbative, long distance physics effects are assumed to factorize into a form factor that must be measured. Default settings and form factors (fit to previous measurements) are used [20]. EV calculates $\gamma^*/Z$ production with full interference, and does the gluon resummation in transverse momentum space. RESBOS calculates $\gamma^*$ or $Z$ boson only production, and does the gluon resummation in the Fourier conjugate impact parameter space. Figure 1 compares the shape of the measured $d\sigma/dP_{\rm T}$ to the $Z$-only RESBOS calculation with CTEQ4M [21] PDFs. For $P_{\rm T} < 80$ GeV/$c$, the EV and RESBOS calculations agree with the measurement, but with deviations of up to 20% that may be due to inaccuracies in the nonperturbative form factor. The precision of the measurement allows for an analysis and refinement of the form factor. For $P_{\rm T} > 80$ GeV/$c$, the calculations agree with the measurement, but with reduced precision. From EV, $\sigma = 231$ pb and 225 pb are predicted with CTEQ4M and MRS-R2 [22] PDFs, respectively. From RESBOS, $\sigma = 233$ pb ($\sigma_{\gamma^*} + \sigma_Z$) with CTEQ4M PDFs. The measurement is $248 \pm 11$ pb.

The measurement of $d\sigma/dP_{\rm T}$ agrees very well with a previous measurement [6] and a concurrent measurement [23]. Total cross section measurements are customarily reported as the $Z$ boson only, $e^+e^-$-pair cross section integrated over all boson masses, $\sigma B(p\bar{p} \to Z \to e^+e^-)$. This derived cross section is found to be $249 \pm 5$ (stat + syst) $\pm$ 10 (lum) pb, where the last error is the luminosity error. This result is consistent with previous measurements [18,24–26], although it is somewhat larger by up to two standard deviations. CDF published [27] the cross section of $\mu^+\mu^-$ pairs from the 1992–1995 data. At publication, the improved luminosity analysis of the 1994–1995 run was not available. With the improvement, $\sigma B(p\bar{p} \to Z \to \mu^+\mu^-) = 237 \pm 9(\text{stat} + \text{syst}) \pm 9(\text{lum})$ pb. The combined $e^+e^-$ and $\mu^+\mu^-$ cross section is $247 \pm 5\,(\text{stat} + \text{syst}) \pm 10\,(\text{lum})$ pb.

In summary, the transverse momentum and total cross sections of Drell-Yan $e^+e^-$ pairs in the mass range 66–116 GeV/$c^2$ produced in $p\bar{p}$ collisions at $\sqrt{s} = 1.8$ TeV have been measured by CDF. Both previous measurements and QCD predictions are consistent with the new measurements. The precision of the $d\sigma/dP_{\rm T}$ measurement allows more of the physics of non-perturbative QCD at low $P_{\rm T}$ to be otained from the data.

The vital contributions of the Fermilab staff and the technical staffs of the participating institutions are gratefully acknowledged. This work is supported by the U.S. Department of Energy and National Science Foundation; the Natural Sciences and Engineering Research Council of Canada; the Istituto Nazionale di Fisica Nucleare of Italy; the Ministry of Education, Science and Culture of Japan; the National Science Council of the Republic of China; the A.P. Sloan Foundation, and the Swiss National Science Foundation.

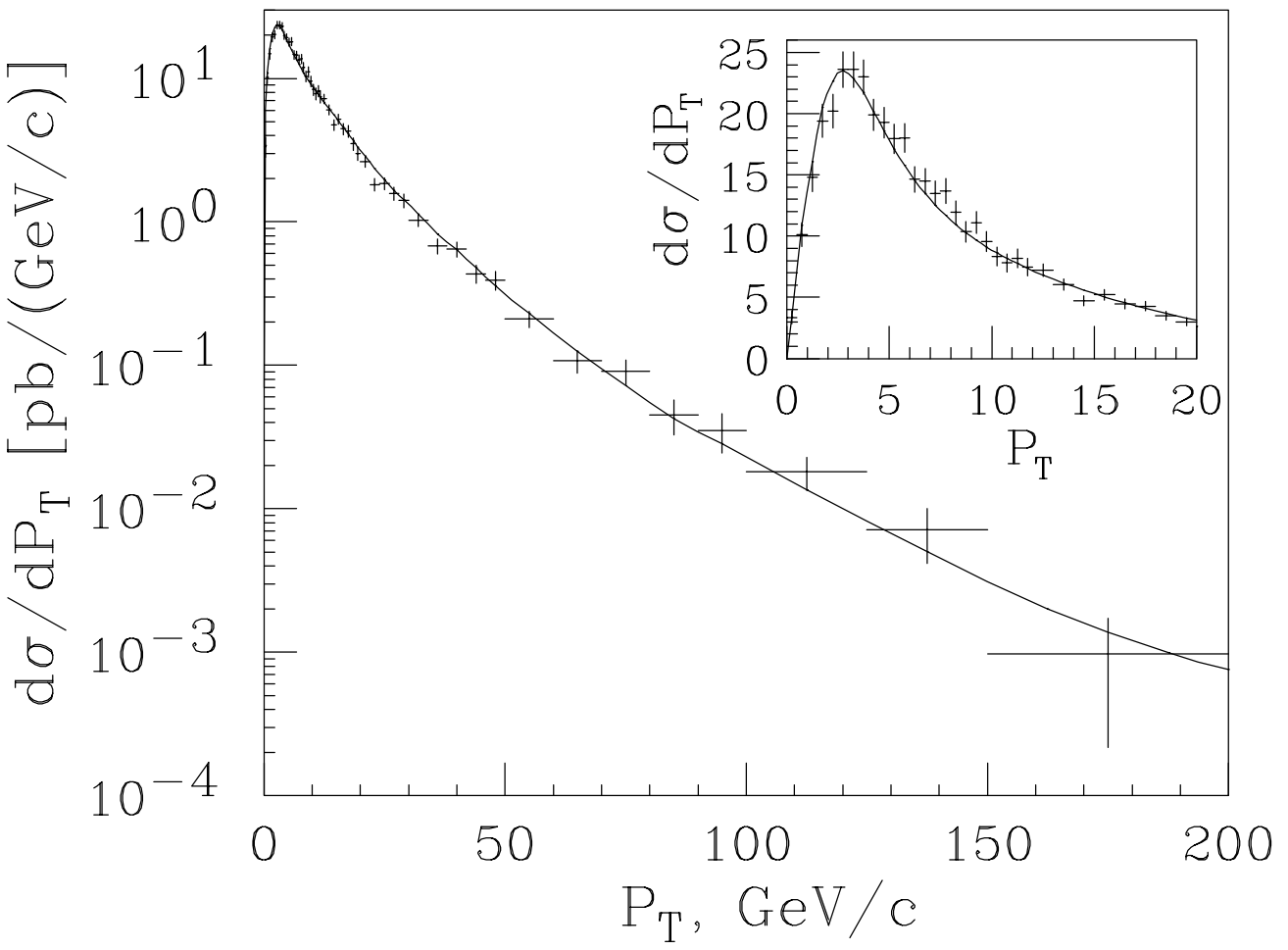

FIG. 1. The $d\sigma/dP_{\mathrm{T}}$ of $e^+e^-$ pairs in the mass range 66–116 GeV/$c^2$. The crosses are the data and the curve is the RESBOS $Z$-only calculation normalized to the data. The insert shows the $P_{\mathrm{T}} < 20$ GeV/$c$ region with a linear ordinate. The data errors do not include the 3.9% luminosity error.

TABLE I. The $d\sigma/dP_{\mathrm{T}}$ of $e^+e^-$ pairs in the mass range 66–116 GeV/$c^2$. All errors are included except the 3.9% integrated luminosity error. The $P_{\mathrm{T}}$ is the bin center value. The bins change size at 0, 12, 20, 30, 50, 100, and 150 GeV/$c$, with corresponding widths of 0.5, 1, 2, 4, 10, 25, and 50 GeV/$c$.

| $P_{\mathrm{T}}$ (GeV/$c$) | $d\sigma/dP_{\mathrm{T}}$ [pb/(GeV/$c$)] | $P_{\mathrm{T}}$ (GeV/$c$) | $d\sigma/dP_{\mathrm{T}}$ [pb/(GeV/$c$)] |
|---|---|---|---|
| 0.25 | $(3.35 \pm 0.54) \times 10^0$ | 13.5 | $(6.05 \pm 0.47) \times 10^0$ |
| 0.75 | $(1.01 \pm 0.10) \times 10^1$ | 14.5 | $(4.73 \pm 0.41) \times 10^0$ |
| 1.25 | $(1.48 \pm 0.12) \times 10^1$ | 15.5 | $(5.21 \pm 0.44) \times 10^0$ |
| 1.75 | $(1.94 \pm 0.14) \times 10^1$ | 16.5 | $(4.46 \pm 0.40) \times 10^0$ |
| 2.25 | $(2.02 \pm 0.14) \times 10^1$ | 17.5 | $(4.28 \pm 0.39) \times 10^0$ |
| 2.75 | $(2.36 \pm 0.15) \times 10^1$ | 18.5 | $(3.51 \pm 0.35) \times 10^0$ |
| 3.25 | $(2.36 \pm 0.14) \times 10^1$ | 19.5 | $(3.01 \pm 0.33) \times 10^0$ |
| 3.75 | $(2.30 \pm 0.14) \times 10^1$ | 21.0 | $(2.63 \pm 0.22) \times 10^0$ |
| 4.25 | $(1.99 \pm 0.13) \times 10^1$ | 23.0 | $(1.82 \pm 0.18) \times 10^0$ |
| 4.75 | $(1.93 \pm 0.12) \times 10^1$ | 25.0 | $(1.85 \pm 0.18) \times 10^0$ |
| 5.25 | $(1.79 \pm 0.12) \times 10^1$ | 27.0 | $(1.58 \pm 0.17) \times 10^0$ |
| 5.75 | $(1.80 \pm 0.12) \times 10^1$ | 29.0 | $(1.41 \pm 0.16) \times 10^0$ |
| 6.25 | $(1.46 \pm 0.10) \times 10^1$ | 32.0 | $(1.02 \pm 0.10) \times 10^0$ |
| 6.75 | $(1.45 \pm 0.10) \times 10^1$ | 36.0 | $(6.78 \pm 0.79) \times 10^{-1}$ |
| 7.25 | $(1.35 \pm 0.10) \times 10^1$ | 40.0 | $(6.44 \pm 0.76) \times 10^{-1}$ |
| 7.75 | $(1.37 \pm 0.10) \times 10^1$ | 44.0 | $(4.34 \pm 0.62) \times 10^{-1}$ |
| 8.25 | $(1.19 \pm 0.09) \times 10^1$ | 48.0 | $(3.94 \pm 0.59) \times 10^{-1}$ |
| 8.75 | $(1.04 \pm 0.09) \times 10^1$ | 55.0 | $(2.10 \pm 0.27) \times 10^{-1}$ |
| 9.25 | $(1.11 \pm 0.09) \times 10^1$ | 65.0 | $(1.07 \pm 0.19) \times 10^{-1}$ |
| 9.75 | $(9.56 \pm 0.82) \times 10^0$ | 75.0 | $(9.10 \pm 1.75) \times 10^{-2}$ |
| 10.25 | $(8.35 \pm 0.76) \times 10^0$ | 85.0 | $(4.50 \pm 1.22) \times 10^{-2}$ |
| 10.75 | $(7.82 \pm 0.74) \times 10^0$ | 95.0 | $(3.51 \pm 1.07) \times 10^{-2}$ |
| 11.25 | $(8.18 \pm 0.76) \times 10^0$ | 112.5 | $(1.81 \pm 0.48) \times 10^{-2}$ |
| 11.75 | $(7.48 \pm 0.72) \times 10^0$ | 137.5 | $(7.11 \pm 2.97) \times 10^{-3}$ |
| 12.50 | $(7.21 \pm 0.53) \times 10^0$ | 175.0 | $(9.74 \pm 7.56) \times 10^{-4}$ |